\def\unit#1{{\hat{#1}}}\def\zhat{\unit z}

\def\inv{^{-1}}
\def\Partial#1#2{{{\partial#1}\over{\partial #2}}}
\def\Func#1#2{{{\delta#1}\over{\delta #2}}}
\def\fr#1#2{{#1\over#2}}
\def\casefr#1#2{\case#1#2}
\def\half{{\casefr12}}

\def\dk{d\vk}

\def\c{\chi}
\def\d{\delta} \def\D{\Delta}
\def\e{\epsilon}

\def\k{\kappa}
\def\Om{\Omega} 
\def\w{\protect\Subcode{\omega}} \def\W{\Omega}

\def\({\left(}\def\){\right)}
\def\[{\left[}\def\]{\right]}
\def\Re{{\rm Re}}

\def\exchange{\leftrightarrow}

\def\Fig#1#2#3{
\begin{figure}
\centerline{\epsffile{#1}}
\caption{#3}\label{#2}
\end{figure}}

\def\It#1{{\it #1}}
\def\Fig#1#2#3{}

\def\Abs#1{{\left|#1\right|}}
\def\cross{{\bbox{\times}}}
\def\vdot{{\bbox{\cdot}}}
\def\v#1{{\bbox{#1}}}

\def\<#1>{\left<#1\right>}
\def\log{\ln}

\def\vk{{\v k}}
\def\vp{{\v p}}
\def\vq{{\v q}}

\def\w{{\omega}}

\def\Eq(#1){Eq.~(\protect\ref{#1})}
\def\Equation(#1){Equation~(\protect\ref{#1})}
\def\eq(#1){\label{#1}}

\def\deltof#1{\Partial{#1}{t}}

\def\kpq{{\vk\vp\vq}}
\def\KPQ{{\vK\vP\vQ}}

\def\nuk{\nu_\vk}

\def\wk{\w_\vk}
\def\wp{\w_\vp}
\def\wq{\w_\vq}

\def\Jkpq{\e_{\kpq}}
\def\DK{\D_\vK} 
\def\DP{\D_\vP}
\def\DQ{\D_\vQ}
\def\vK{{\v K}}
\def\vP{{\v P}}
\def\vQ{{\v Q}}
\def\WK{\W_\vK}
\def\WP{\W_\vP}
\def\WQ{\W_\vQ}
\def\cD{{\cal D}}

\documentstyle[prl,aps,amsfonts,preprint]{revtex}

\input{epsf.sty}

\begin{document}

\bibliographystyle{pfb}

\title{Spectral reduction: a statistical description of turbulence}

\author{John C. Bowman,$^1$ B. A. Shadwick,$^{2,3}$ and P. J. Morrison$^4$}
\address{$^1$Department of Mathematical Sciences, University of Alberta,
Edmonton, Alberta, Canada T6G~2G1}
\address{$^2$The Institute for Advanced Physics, Conifer C0 80433--7727}
\address{$^3$Physics Department, University of California at Berkeley,
Berkeley CA 94720--7300}
\address{$^4$Department of Physics and Institute for Fusion Studies, The
University of Texas at Austin, Austin TX 78712--1081}

\date{\today}

\maketitle

\begin{abstract}

A method is described for predicting statistical properties of
turbulence. Collections of Fourier amplitudes are represented by
nonuniformly spaced modes with enhanced coupling coefficients. The
statistics of the full dynamics can be recovered from the time-averaged
predictions of the reduced model. A Liouville theorem leads to
inviscid equipartition solutions. Excellent agreement is obtained with
two-dimensional forced-dissipative pseudospectral simulations. For the
two-dimensional enstrophy cascade, logarithmic corrections to the high-order
structure functions are observed.

\end{abstract}


\narrowtext

Many practical applications for spectral simulations of turbulence exist
where it would be desirable to evolve modes that are distributed
nonuniformly in Fourier space, devoting most of the computational resources
to the length scales of greatest physical interest. This idea has led to the
development of a new reduced statistical description of turbulence, called
{\it spectral reduction\/}\cite{Bowman95d}, which dramatically
reduces the number of spectral modes that must be retained in simulations of
turbulent phenomena. It exploits the fact that statistical moments are much
smoother functions of wave number than are the underlying stochastic
amplitudes.

The concept of wave-number reduction is not new. In the method of
constrained decimation \cite{Kraichnan85,Williams87,Kraichnan89}, a
stochastic forcing term is added to model the effect of the deleted modes
on the retained modes.  She and Jackson have proposed a reduction scheme in
which the linear (viscous) term is modified\cite{She93}. In spectral
reduction, a third alternative is chosen: the nonlinear coefficients are
enhanced to account for the effect of the discarded modes on the explicitly
evolved modes.  There have been other more heuristic attempts at
wave-number reduction\cite{Lorenz72,Lee89,Vazquez-Semadeni92,Grossmann92};
these methods typically neglect nonlocal wave-number triad interactions
(which play a particularly important role in two-dimensional
turbulence). Unlike the renormalization group \cite{Yakhot86} method, which
retains only large-scale modes and attempts to express the effect of the
small-scale modes using a self-similarity {\it Ansatz\/}, spectral
reduction retains certain modes from all scales, while discarding other
modes associated with these same scales. The generality of the formulation
allows one to refine the partition wherever the physics dictates.

In this Letter we restrict our attention to homogeneous and isotropic
incompressible turbulence in two dimensions.  The appropriate spectral
transform in this limit is the integral Fourier transform, under which the
two-dimensional Navier--Stokes vorticity equation becomes
\begin{equation}
\deltof{\wk}+\nuk\wk =\int_\cD d\vp\int_\cD d\vq\,
\fr{\Jkpq}{q^2} \wp^*\wq^*.
\eq(continuum)
\end{equation}
Here $\nuk$ models time-independent linear dissipation or forcing and the
interaction coefficient
$\Jkpq\doteq(\zhat\vdot\vp\cross\vq)\,\d(\vk+\vp+\vq)$ is antisymmetric
under permutation of any two indices ($*$ denotes complex conjugation and
$\doteq$ indicates a definition).
We restrict the integration to a bounded wave-number domain $\cD$ that
excludes a neighborhood of $\vk=\v0$.  As a consequence of the antisymmetry
of $\Jkpq$, in the inviscid limit ($\nuk=0$) \Eq(continuum) conserves the
energy~$\half\int_\cD d\vk\,\Abs{\wk}^2/k^2$ and
enstrophy~$\half\int_\cD d\vk\,\Abs{\wk}^2$. It is believed that the energy
and enstrophy play fundamental roles in the dynamics of the turbulent cascade.

We introduce an arbitrary coarse-grained grid that partitions
$\cD$ into connected regions called bins. The bins are labeled by capital
letters to distinguish them from the
continuum wave numbers, which we represent by lower-case letters.
To this grid, we associate new variables
$\WK\doteq\DK\inv\int_{\DK}\wk\,d\vk,$ where $\DK$ is the area of bin~$\vK$.
The exact evolution of $\WK$ is given by
\begin{equation}
\deltof{\WK}+\<\nuk\wk>_\vK =
\sum_{\vP,\vQ} \DP \DQ\<\fr{\Jkpq}{q^2}\wp^*\wq^*>_{\KPQ},\eq(binaveraged)
\end{equation}
where $\<\cdot>_\vK$ denotes a bin average and the operator
\begin{equation}
\<f>_{\KPQ}\doteq\fr{1}{\DK\DP\DQ}\int_{\DK}\dk \int_{\DP} d\vp\int_{\DQ} d\vq
\,f,\eq(favg)
\end{equation}
depends only on the bin geometry.
The geometric factors $\<f>_{\KPQ}$ can be efficiently
computed using a combination of analytical and numerical
methods\cite{Bowman92,Bowman95b,Bowman97}. Since they are independent of
both time and initial conditions, they need only be computed once for each
new wave-number partition. The reality condition $\WK=\Om_{-\vK}^*$,
where $-\vK$ denotes the inversion of bin $\vK$ through the origin, will be
respected for partitions that possess inversion symmetry.

\Equation(binaveraged) is unfortunately not closed. If $\wk$ were
naively approximated by its bin-averaged value $\WK$, one would obtain
\begin{equation}
\deltof\WK+\<\nuk>_\vK\WK=\sum_{\vP,\vQ} \DP \DQ
\<\fr{\Jkpq}{q^2}>_{\KPQ}\WP^*\WQ^*.\eq(closure)
\end{equation}
In the inviscid limit, \Eq(closure) conserves the coarse-grained enstrophy
$\half\sum_\vK \Abs{\WK}^2\DK$ since $\<\Jkpq/q^2>_{\KPQ}$ is antisymmetric in
$\vK\exchange\vP$. However, the coarse-grained energy $\half\sum_\vK
\Abs{\WK}^2\DK/K^2$ is not conserved since $\<\Jkpq/q^2>_{\KPQ}/K^2$ is not
antisymmetric in $\vK\exchange\vQ$ (here $K$ denotes the magnitude of some
characteristic wave number in bin $\vK$). However, both of these desired
symmetries can be reinstated by replacing the factor $\<\Jkpq/q^2>_\KPQ$
in~\Eq(closure) with the slightly modified coefficient
$\<\Jkpq>_\KPQ/Q^2$. The relative error introduced by this modification
is negligible in the limit of small bin size, being on the order of the
squared relative variation in the wavenumber magnitude over a bin. The
result,
\begin{equation}
 \deltof\WK+\<\nuk>_\vK\WK
=\sum_{\vP,\vQ} \DP \DQ \fr{\<\Jkpq>_{\KPQ}}{Q^2} \WP^*\WQ^*,
\eq(SR)
\end{equation}
which we call the spectrally reduced Navier--Stokes equation, is a more
acceptable alternative than~\Eq(closure) as a closure of~\Eq(binaveraged):
not only does it reduce to the Navier--Stokes equation in this limit, but it
also conserves both energy and enstrophy, even when the bins are large. The
final modification leading to~\Eq(SR) partially compensates for the error
introduced by the crude approximation $\wk\approx\WK$ and leads to the same
general structure and symmetries as~\Eq(continuum); in this sense spectral
reduction may be regarded as a renormalization.

If the bins are large, the true vorticity will vary rapidly with wave number
within each bin and it is unlikely that~\Eq(SR) will yield a reasonable
description of the instantaneous dynamics. However, the time-averaged (or
ensemble-averaged) moments of~\Eq(SR) satisfy equations that closely
approximate the equations governing the exact bin-averaged statistics.
For example, a time average (denoted by an over-bar) of the bin-averaged
enstrophy equation derived from~\Eq(continuum) leads to
\begin{eqnarray}
&&\fr{1}{2}\overline{\deltof{\<\Abs{\wk}^2>_\vK}}+
\Re\<\nuk\overline{\Abs{\wk}^2}>_\vK\
\nonumber\\
&=&\Re\sum_{\vP,\vQ}\DP\DQ\<\fr{\Jkpq}{q^2}\overline{\wk^*\wp^*\wq^*}>_{\KPQ}.
\eq(bintimeaveraged) 
\end{eqnarray}
If the true vorticity is a continuous function of wave number, the mean
value theorem for integrals guarantees the existence of a wave number $\v\k$
in bin $\vK$ such that $\WK=\w_{\v\k}$.  Furthermore, time-averaged
quantities such as $\overline{\Abs{\wk}^2}$ are generally smooth functions
of the wave number $\vk$.  We thus deduce that
$\overline{\Abs{\WK}^2}=\overline{\Abs{\w_{\v\k}}^2}\approx
\overline{\Abs{\wk}^2}$ for all $\vk$ in bin $\vK.$
Similarly, the triplet correlation $\overline{\wk^*\wp^*\wq^*}$ is a smooth
function of $\vk$, $\vp$, $\vq$ when restricted to the surface defined by
the triad condition $\vk+\vp+\vq=\v0$.

To good accuracy the statistical averages in~\Eq(bintimeaveraged) may
therefore be evaluated at the characteristic wave numbers $\vK$, $\vP$,
$\vQ$ of each bin, yielding
\begin{eqnarray}
\fr{1}{2}\overline{\deltof{\Abs{\WK}^2}}&+&\Re\<\nuk>_\vK\overline{\Abs{\WK}^2}
\nonumber\\
&=&\Re\sum_{\vP,\vQ} \DP \DQ\<\fr{\Jkpq}{q^2}>_{\KPQ} \overline{\WK^*\WP^*\WQ^*}.
\end{eqnarray}
Moreover, to the extent that the wave-number magnitudes vary slowly over a
bin,~\Eq(bintimeaveraged) may equally well be reduced to the (nonlinearly
conservative) approximation
\begin{eqnarray}
\fr{1}{2}\overline{\deltof{\Abs{\WK}^2}}&+&\Re\<\nuk>_\vK\overline{\Abs{\WK}^2}
\nonumber\\
&=&\Re\sum_{\vP,\vQ} \DP \DQ \fr{\<\Jkpq>_{\KPQ}}{Q^2}
\overline{\WK^*\WP^*\WQ^*},\eq(SRtimeaveraged)
\end{eqnarray}
which is precisely the evolution equation for the time-averaged enstrophy
obtained from~\Eq(SR). Similar arguments for the higher-order statistical
moments can also be made, suggesting that spectral reduction can indeed
provide an accurate statistical description of turbulence, even when each
bin contains many statistically independent modes.  As the
partition is refined, one expects the solutions of~\Eq(SRtimeaveraged)
to converge to the those of~\Eq(bintimeaveraged).

In the absence of forcing and dissipation, the (untruncated) two-dimensional
Euler equations can be written in a noncanonical Hamiltonian
framework \cite{Morrison97} as $\dot\wk=\int d\vq\, J_{\vk\vq} \d H/\d
\wq$, where $H\doteq\half\int d\vk\,\Abs{\wk}^2/k^2$ and
$J_{\vk\vq}\doteq\int d\vp\,\Jkpq \wp^*$. The Liouville theorem
$$
\int d\vk\Func{\dot\wk}{\wk}=
\int d\vk\int d\vq\left[
\e_{\vk(-\vk)\vq}\Func{H}{\wq}+J_{\vk\vq}\fr{\d^2 H}{\d \wk\d\wq}\right]=0
$$
then follows immediately from the properties of $\Jkpq$.

When $\nuk=0$, \Eq(SR) can be written in a similar form as
$\dot\WK=\sum_\vQ J_{\vK\vQ}\,\partial H/\partial\WQ$, where $H\doteq
\half\sum_\vK \Abs{\WK}^2 \DK/K^2$ and
\begin{equation}
J_{\vK\vQ}\doteq\sum_\vP\DP \<\Jkpq>_{\KPQ}
\WP^*.
\eq(SR-J)
\end{equation}
It is an open question whether (an untruncated version of)~\Eq(SR-J)
satisfies the Jacobi identity, which would make spectral reduction a
Hamiltonian approximation. What is certain is that the respective Liouville
theorem
$$
\sum_\vK\Partial{\dot\WK}{\WK}
=\sum_{\vK,\vQ}\(\Partial{J_{\vK\vQ}}{\WK}\Partial{H}{\WQ}+
J_{\vK\vQ}\fr{\partial^2 H}{\partial \WK\partial\WQ}\)=0
$$
is obeyed, as a consequence of the antisymmetry of
$\<\Jkpq>_{\vK\vP\vQ}$. If the dynamics is \It{mixing}, the inviscid system
will then evolve toward equipartition\cite{Orszag70,Carnevale81}; this was
verified for spectral reduction numerically, using a fifth-order
conservative Runga--Kutta integration algorithm that conserves quadratic
invariants to all orders in the time step \cite{Bowman97b,Shadwick99}. 
When using nonuniform bins it is necessary to rescale the time derivative
$\partial/\partial t$ in \Eq(SR) to $(\D_0/\DK)\partial/\partial t$, where
$\D_0$ is the minimum bin area, to obtain an equipartition of modal
(instead of bin-averaged) energies. This modification to the transient
evolution will be discussed further in a future paper.

Upon adding to \Eq(SR) a random stirring force for $k\in [5,7]$ and adopting
the dissipation function $\nuk=\nu_L \theta(7-k)+\nu_H k^2$ ($\theta$ is
the Heaviside function), we graph in Fig.~\ref{ekvkconvo} the
time-averaged saturated energy spectra for four wave-number partitions
to test how rapidly spectral reduction converges.
The excellent agreement demonstrated
between the predictions of spectral reduction and a (computationally more
expensive) full dealiased pseudospectral simulation is obtained without
fitting or the introduction of adjustable parameters. We also show
the predictions of the realizable test field model
(RTFM)\cite{Bowman97}, using wavenumber binning and setting the
eddy-damping multiplier in this heuristic statistical closure to one.

\Fig{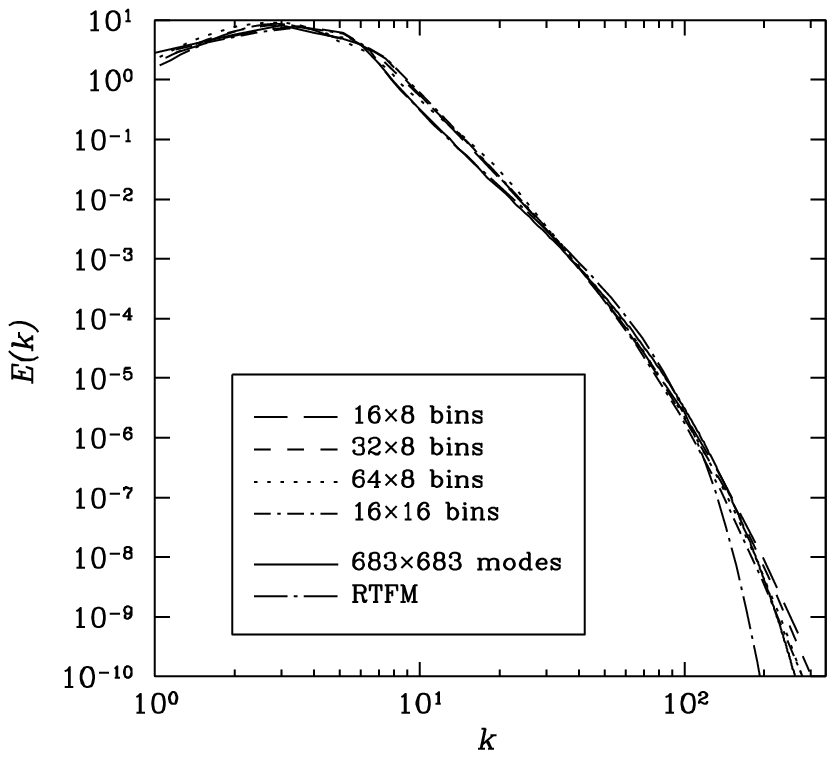}{ekvkconvo}{Comparison of the turbulent energy
spectra obtained with $16\times 8$, $32\times 8$, $64\times 8$, and
$16\times 16$ (logarithmically spaced radial~$\times$ uniformly spaced angular)
wave-number partitions, the RTFM, and a full~$683\times 683$ dealiased
pseudospectral simulation ($1024\times 1024$ total modes).}

High-order moments are also accurately described by spectral reduction.  A
quantity of interest is the angular average $S_n(r)$ of the~$n$-th
(time-averaged) moment of velocity increments~$\overline{\Abs{v(\v
r)-v(\v0)}^n}$, or \It{structure function}.  In Fig.~\ref{structconv10o} we
illustrate the scaling with distance $r$ of a typical high-order structure
function, $S_{10}(r)$, for the runs depicted in
Fig.~\ref{ekvkconvo}. Slight variations in the predicted large-scale
velocities are evident as overall vertical offsets. Note that the
(discrete) pseudospectral calculation is an approximation to \Eq(continuum)
at the large scales. 

\Fig{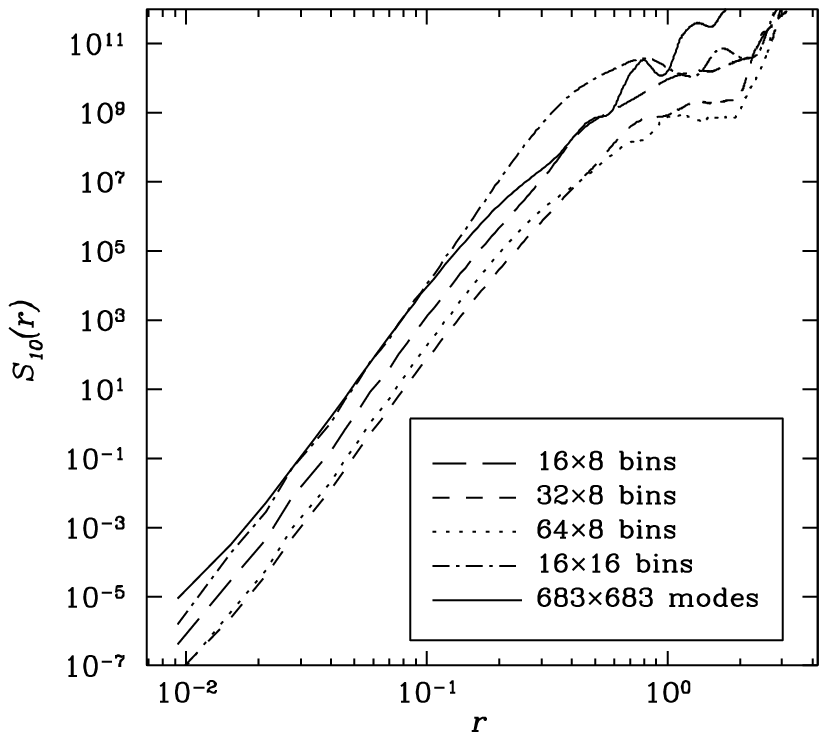}{structconv10o}{Angle-averaged structure function $S_{10}(r)$.} 

One can readily investigate high-Reynolds number turbulence with spectral
reduction, using a polar partition in which the bins are logarithmically
spaced in the radial wave number. A saturated turbulent state can be
evolved for thousands of eddy turnover times to obtain statistically
meaningful moments for comparison with theoretical predictions. For
example, Kolmogorov's idea of self-similar energy transfer in the inertial
range\cite{Kolmogorov41} led Kraichnan\cite{Kraichnan71b} to propose a
logarithmically corrected asymptotic form for the energy spectrum $E(k)$ of
the enstrophy cascade.  In a simulation with viscous dissipation active
only at the smallest scales (to yield a pristine inertial range) and
forcing via a linear instability (negative $\nuk$), we apply spectral
reduction to demonstrate the recent extension $E(k)\sim k^{-3}\c^{-1/3}(k)$
of Kraichnan's result to the entire inertial range, where $k_1$ is the
smallest inertial-range wave number, $\c(k)\doteq \log(k/k_1)+\c_1$, and
the positive constant $\c_1$ is set by the large-scale
dynamics\cite{Bowman96}.  We verify in Fig.~\ref{32x8ailogkchi1} the linear
behavior of $[k^3 E(k)]^{-3}$ with respect to $\log(k/k_1)$, using the
values $k_1=16.4$ and $\c_1=0.67$ determined by a least-squares fit. The
inertial-range energy spectrum is thus well described by Kraichnan's
logarithmically corrected $k^{-3}$ law.

For the second simulation, we demonstrate in Fig.~\ref{32x8alogrchi1comp} the
linear behavior of $[r^{-n} S_n(r)]^{3/n}$ with $\log(r_1/r)$ on the
interval $0.043 \le r\le r_1=0.26$ for various values of $n$. The implied
scaling $S_n(r)\sim r^n [\log (r_1/r)+\chi_n']^{n/3}$, where $\chi_n'$
is a constant, is in agreement with both the asymptotic theory of Falkovich
and Lebedev\cite{Falkovich94} and the recent experimental results of
Paret \It{et al.}\cite{Paret99}, lending support to the claim that there are no
high-order intermittency corrections in two-dimensional turbulence. The
universality of this result will be investigated in a future paper.

\Fig{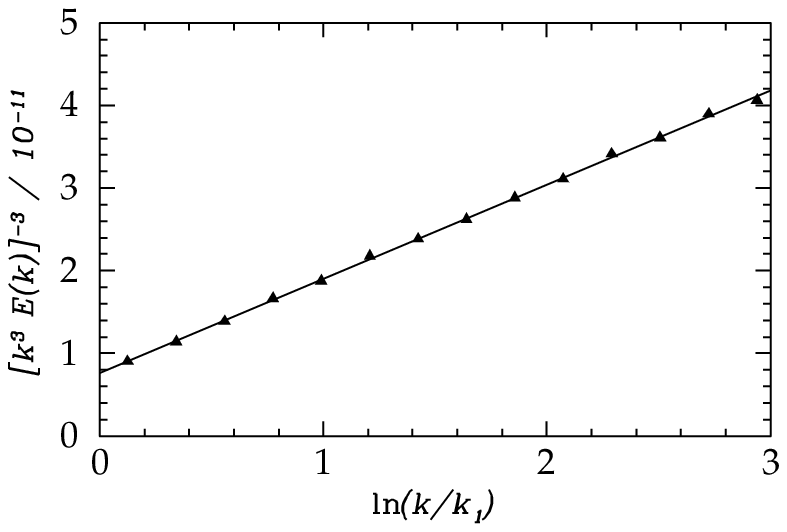}{32x8ailogkchi1}{Linearity of $[k^3 E(k)]^{-3}$ with
respect to $\log(k/k_1)$ for an enstrophy inertial range between $k_1=16.4$
and $k=330$. The solid triangles are the predictions of spectral
reduction.}

\Fig{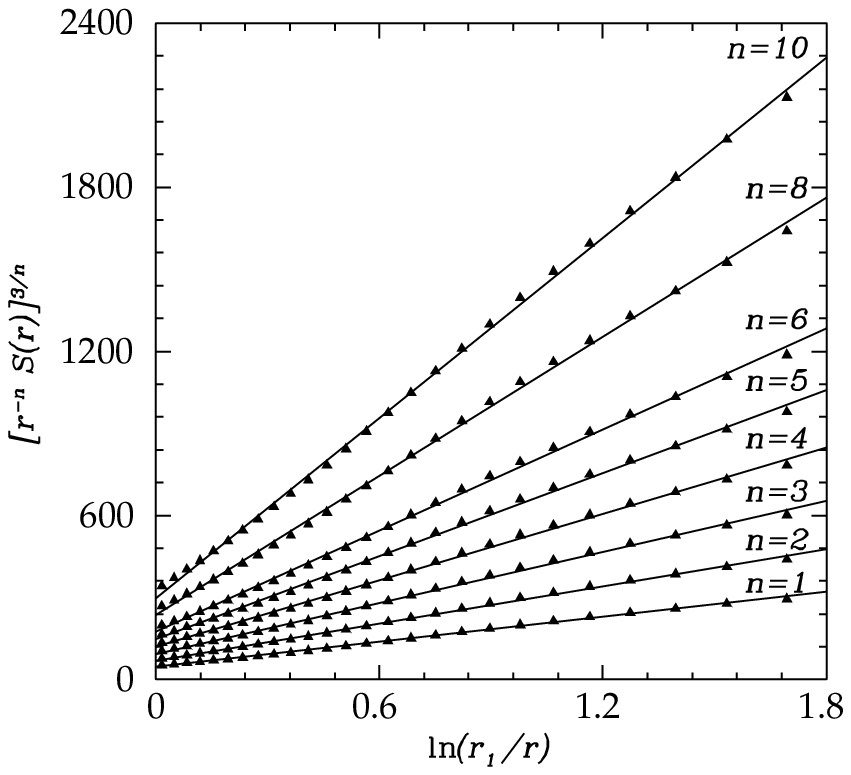}{32x8alogrchi1comp}{Linearity of $[r^{-n} S(r)]^{3/n}$
with respect to $\log(r_1/r)$ for $0.043\le r \le r_1=0.26$.}

In this Letter, we propose a new technique that dramatically
decreases the number of degrees of freedom required to simulate homogeneous
turbulence. The statistically stationary state described by
Fig.~\ref{32x8ailogkchi1}, which would require $2048\times 2048$ dealiased
($3071\times3071$ total) pseudospectral modes, can be successfully modeled
using only $32\times8$ bins. A notable feature of spectral reduction
that distinguishes it from other statistical theories of turbulence is the
existence of a control parameter (bin size) that can be varied to increase
the accuracy of a solution.  Moreover, spectral reduction does not make a
closure assumption on the triplet correlation $\overline{\WK^*\WP^*\WQ^*}$
appearing in~\Eq(SRtimeaveraged); it circumvents the closure problem
entirely by reducing the number of triplet correlations to a tractable
number, instead of eliminating them in favor of lower-order statistical
variables. Unlike statistical closures, spectral reduction thus does not
destroy the phase information embodied in the triplet correlation.

Spectral reduction appears to be a promising candidate as a statistical
description of turbulence. We propose that it could be used to assess the
effect of various dissipation mechanisms in large-eddy simulations, as a
subgrid model, or even as a substitute for full simulation of high-Reynolds
number turbulence. However, as it does not provide explicit insight into
underlying dynamical processes, spectral reduction should be considered
more as a computational tool than as a true analytical theory of
turbulence. The latter challenge still awaits us.

We would like to thank John A. Krommes, for the idea of anisotropic
bin-averaging of statistical closure equations, R. E. Waltz, for the
suggestion of applying that technique directly to the Navier--Stokes
equations, and C. E. Leith for discussions on the Liouville theorem.
This work was supported by U.S. Department of Energy Contract
No.\ DE--FG03--96ER--54346 and the Natural Sciences and Engineering
Research Council of Canada.

\def\Fig#1#2#3{
\begin{figure}[h]
\centerline{\epsffile{#1}}
\caption{#3}\label{#2}
\end{figure}}

\Fig{fig1.eps}{ekvkconvo}{Comparison of the turbulent energy
spectra obtained with $16\times 8$, $32\times 8$, $64\times 8$, and
$16\times 16$ (logarithmically spaced radial~$\times$ uniformly spaced angular)
wave-number partitions, the RTFM, and a full~$683\times 683$ dealiased
pseudospectral simulation ($1024\times 1024$ total modes).}

\Fig{fig2.eps}{structconv10o}{Angle-averaged structure function $S_{10}(r)$.} 

\Fig{fig3.eps}{32x8ailogkchi1}{Linearity of $[k^3 E(k)]^{-3}$ with
respect to $\log(k/k_1)$ for an enstrophy inertial range between $k_1=16.4$
and $k=330$. The solid triangles are the predictions of spectral
reduction.}

\Fig{fig4.eps}{32x8alogrchi1comp}{Linearity of $[r^{-n} S(r)]^{3/n}$
with respect to $\log(r_1/r)$ for $0.043\le r \le r_1=0.26$.}

\end{document}